\documentstyle[12pt] {article}
\newcommand{\beq}{\begin{equation}}
\newcommand{\eeq}{\end{equation}}

\begin{document}

\begin{center}
{\Large \bf Ideal Quantum Gases in D-dimensional Space 
and Power-Law Potentials} 
\vskip 0.8cm 
Luca Salasnich \\
\vskip 0.5cm
Istituto Nazionale per la Fisica della Materia, Unit\`a di Milano,\\ 
Dipartimento di Fisica, Universit\`a di Milano, \\ 
Via Celoria 16, 20133 Milano, Italy \\
e-mail: salasnich@mi.infm.it 
\end{center}

\vskip 1.5cm

\begin{center}
{\bf Abstract}
\end{center} 
We investigate ideal quantum gases in D-dimensional space and confined 
in a generic 
external potential by using the semiclassical approximation. 
In particular, we derive density of states, 
density profiles and critical temperatures 
for Fermions and Bosons trapped in isotropic power-law potentials. 
Form such results, one can easily obtain those of quantum gases 
in a rigid box and in a harmonic trap. 
Finally, we show that the Bose-Einstein condensation 
can set up in a confining power-law potential 
if and only if ${D/2}+{D/n}>1$, 
where $D$ is the space dimension and $n$ is the 
power-law exponent. 

\vskip 0.5cm


\newpage 

\section{Introduction}

For dilute alkali-metal atoms in magnetic or 
magneto-optical traps at very low temperatures, 
the Bose-Einstein condensation has been achieved$^{1}$ in 1995
and the Fermi quantum degeneracy$^{2}$ in 1999. 
These results have renewed the theoretical investigation 
on Bose and Fermi gases. 
\par 
In the experiments with Bosons, the system is weakly-interacting 
and the thermodynamical properties depend on the s-wave scattering length 
(for a review see Ref. 3). Nevertheless, by using Feshback resonances, 
it is now possible to modify and also switch-off 
the atom-atom interaction.$^{4}$ In the case of Fermions, 
the s-wave scattering between atoms in the same hyperfine state 
is inhibited due the Pauli principle. It follows that at low temperature 
the dilute Fermi gas, in a fixed hyperfine state, is practically ideal.$^{2}$ 
\par 
In previous papers we analyzed ground-state and vortex 
properties of Bose condensates 
in different external potentials: 
harmonic potential,$^{5-10}$, toroidal potential$^{11}$ 
and double-well potential.$^{12}$ Recently, 
we have also afforded the study of the thermodynamics 
of interacting Bose gases in harmonic potential.$^{13,14}$ 
\par  
In this paper, we investigate the thermal 
properties of both Bose and Fermi ideal gases in a 
generic confining external potential.  
All the calculations are performed by 
assuming a D-dimensional space. Such an assumption 
is motivated by esthetic criteria but also 
by recent experiments with degenerate gases 
in systems with reduced or fractal dimension.$^{3}$ 
We analyze in detail the isotropic power-law potential, 
from which one easily deduces the results of a rigid box 
and a harmonic trap. 

\section{Confined Ideal Fermi and Bose Gases}

Let us consider a confined quantum gas of non-interacting identical 
Fermions (or Bosons) in D-dimensional space. 
In the grand canonical ensemble of equilibrium 
statistical mechanics,$^{15}$  
the average number $N_{\alpha}$ of particles in the single-particle 
state $|\alpha\rangle$ with energy $\epsilon_{\alpha}$ is given by 
\beq
N_{\alpha} = {1\over e^{\beta(\epsilon_{\alpha}-\mu)} \pm 1}  \; , 
\eeq
where the sign $+$ ($-$) is for Fermions (Bosons), 
$\mu$ is the chemical potential and $\beta=1/(kT)$ 
with $k$ the Boltzmann constant and $T$ the absolute temperature. 
In general, given the single-particle function $N_{\alpha}$, 
the average total number $N$ of particles of the system reads 
\beq
N= \sum_{\alpha} N_{\alpha}  \; . 
\eeq
This condition fixes the chemical potential $\mu$. 
Thus, $\mu$ is a function of $\beta$ and $N$. 
In the case of Fermions, $\mu$ has no limitations 
and at zero temperature $\mu$ is called Fermi energy $E_F$. 
From the Fermi energy $E_F$ one 
immediately obtains the Fermi temperature $T_F=E_F/k$. 
Below the Fermi temperature, the Fermions begin to fill 
the lowest available single-particle states 
in accordance with the Pauli exclusion Principle: 
one has the Fermi quantum degeneracy.$^{15}$ 
In the case of Bosons, 
$\mu$ cannot be higher than the lowest single-particle energy level 
$\epsilon_0$, i.e., it must be $\mu < \epsilon_0$. 
When $\mu \to \epsilon_0$ the function $N_0$ diverges 
and consequently also $N$ diverges. The physical meaning is that the lowest 
single-particle state becomes macroscopically occupied and 
one has the so-called Bose-Einstein condensation (BEC).$^{15}$ 
It is a standard procedure to calculate the condensed 
fraction $N_0/N$ and also the BEC transition temperature $T_B$ 
by studying the non divergent quantity $N-N_0$ at $\mu =\epsilon_0$ 
as a function of the temperature.$^{3,15}$  
\par 
In the semiclassical limit, the D-dimensional 
system is described by a continuum 
of states$^{15-17}$ and, instead of $\epsilon_{\alpha}$, 
one uses the classical single-particle phase-space energy 
$\epsilon({\bf r},{\bf p})$, where ${\bf r}=(r_1,...,r_D)$ 
is the position vector and ${\bf p}=(p_1,...,p_D)$ 
is the linear momentum vector. 
In this way one obtains from Eq. (1) 
the single-particle phase-space distribution 
\beq 
n({\bf r},{\bf p}) = 
{1\over e^{\beta(\epsilon({\bf r},{\bf p})-\mu)} \pm 1}  \; . 
\eeq 
Note that the accuracy of the semiclassical approximation is expected 
to be good if the number of particles is large 
and the energy level spacing is smaller then $kT$.$^{15-17}$ 
Because of the Heisenberg principle, 
the quantum elementary volume of the single-particle 
2D-dimensional phase-space is 
given by $(2\pi\hbar)^D$, where $\hbar$ is the Planck constant.$^{16}$ 
It follows that the average number $N$ of particles 
in the D-dimensional space can be written as 
\beq
N= \int {d^D{\bf r} \; d^D{\bf p} \over (2\pi \hbar )^D}
\; n({\bf r},{\bf p}) =
\int d^D{\bf r} \; n({\bf r})=\int d^D{\bf p} \; n({\bf p}) \; ,
\eeq
where 
\beq
n({\bf r})=\int {d^D{\bf p} \over (2\pi \hbar )^D}
n({\bf r},{\bf p}) \; 
\eeq
is the spatial distribution, and 
\beq
n({\bf p})=\int {d^D{\bf r} \over (2\pi \hbar )^D}
n({\bf r},{\bf p}) \; 
\eeq
is the momentum distribution. It is important to observe that 
the total number $N$ of particle can also be written as 
\beq
N=\int_0^{\infty} 
d\epsilon \; \rho(\epsilon ) \; 
{1\over e^{\beta(\epsilon - \mu)} \pm 1} \; , 
\eeq 
where $\rho (\epsilon )$ is the density of states. 
It can be obtained from the semiclassical formula 
\beq
\rho(\epsilon ) = \int {d^D{\bf r} \; d^D{\bf p}\over (2\pi\hbar)^D} 
\delta (\epsilon - \epsilon({\bf p},{\bf r})) \; , 
\eeq 
where $\delta(x)$ is the Dirac delta function. 
\par
In the case of Fermions, at zero temperature, i.e., 
in the limit $\beta\to \infty$ where 
$\mu\to E_F$ (the Fermi energy), the phase-space distribution (3) 
becomes 
\beq 
n({\bf r},{\bf p}) = \Theta
\left(E_F - \epsilon({\bf r},{\bf p}) \right) \; , 
\eeq 
where $\Theta(x)$ is the Heaviside step function.$^{15,16}$  
\par
In the case of Bosons, below the BEC transition temperature $T_B$, 
the equation (3) describes only the non-condensed thermal cloud. 
Thus, the semiclassical 
quantization renormalizes the exact Bose distribution (1) that is 
divergent. Because there is not an unique way to translate wave-functions 
into a phase-space distribution,$^{16}$ one cannot introduce an 
exact single-particle phase-space distribution for the Bose condensate. 
Nevertheless, the exact spatial distribution of the 
Bose condensate is $n_0({\bf r})=|\Psi({\bf r})|^2$, 
where $\Psi({\bf r})$ is called order parameter or macroscopic 
wave-function of the condensate, normalized to the number $N_0$ of 
condensed Bosons. For an ideal Bose gas, the function 
$\Psi({\bf r})$ is simply the eigenfunction of the lowest 
single-particle state of the system. 
In this paper we do not study 
the density profiles of the Bose condensed fraction because, 
for a non-interacting gas, their shape is not temperature 
dependent: only their normalization is a function 
of temperature. Actually, to calculate the BEC transition temperature $T_B$ 
and the condensed fraction $N_0/N$ it is sufficient to study 
the non-condensed fraction (thermal cloud). 
(For a recent discussion of the properties of the Bose 
condensate, see Ref. 3 and also Ref. 5-14). 

\section{Gases in External Potential}

Let us consider the ideal Fermi (Bose) gas in a confining external 
potential $U({\bf r})$ and in a D-dimensional space. 
The classical single-particle energy is defined as 
\beq 
\epsilon({\bf r},{\bf p})={{\bf p}^2\over 2m} + U({\bf r}) \; ,
\eeq 
where ${\bf p}^2/(2m)$ is the kinetic energy and 
$m$ is the mass of the particle. The D-dimensional 
vectors ${\bf r}=(x_1,...,x_D)$ and ${\bf p}=(p_1,...,p_D)$ 
are respectively the position and momentum of the particle. 
\par
First, we note that, by using Eq. (8) and Eq. (10), 
the semiclassical density of states can be written as 
\beq 
\rho(\epsilon ) = 
\left({m\over 2\pi \hbar^2}\right)^{D\over 2} 
{1\over \Gamma({D\over 2})} 
\int d^D{\bf r} \; 
\left( \epsilon - U({\bf r}) \right)^{(D-2)\over 2} 
\; . 
\eeq 
where $\Gamma(n)$ is the factorial function. 
Then, we introduce the Fermi and Bose functions.$^{18,19}$  

\vskip 0.5cm

\par 
DEFINITION 1. {\it The Fermi function is given by 
$$ 
f_{n}(z)={1\over \Gamma(n)}\int_0^{\infty} dy 
{z e^{-y} y^{n-1} \over 1 + z e^{-y} } \; ,
$$
and the Bose function is 
$$ 
g_{n}(z)= {1\over \Gamma(n)}\int_0^{\infty} dy 
{z e^{-y} y^{n-1} \over 1 - z e^{-y} } \; ,
$$
where $\Gamma(n)$ is the factorial function.}

\vskip 0.5cm

\par 
REMARK 1. {\it The Fermi and Bose functions are connected by 
the relation $f_n(z)=-g_n(-z)$. For $|z|<1$ the two 
functions can be written as 
$$
g_n(z)=\sum_{i=1}^{\infty} {z^i\over i^{n} } \; , 
$$ 
$$
f_n(z)=\sum_{i=1}^{\infty} (-1)^{i+1} {z^i \over i^{n} } \; .  
$$
One finds also that $g_n(1)=\zeta(n)$, where $\zeta(n)$ is the 
Riemann $\zeta$-function.$^{18,19}$}

\vskip 0.5cm

\par
Now we state two theorems about ideal Fermi and Bose gases 
in external potential. Remind that we work in 
the semiclassical limit.$^{3,15-16}$ 

\vskip 0.5cm

\par
THEOREM 1.1 {\it For an ideal Fermi gas in an external 
potential $U({\bf r})$ and D-dimensional space, 
the finite temperature spatial distribution 
is given by  
$$
n({\bf r})={1\over \lambda^D} f_{D\over 2}
\left(e^{\beta(\mu -U({\bf r}))}\right) \; ,
$$
where $\lambda = (2\pi \hbar^2\beta /m)^{1/2}$ is the 
thermal length and $\mu$ is the chemical potential. 
The zero temperature spatial distribution is 
$$ 
n({\bf r})=\left({m\over 2\pi \hbar^2}\right)^{D/2} 
{1\over \Gamma({D\over 2}+1)} 
\left(E_F- U({\bf r})\right)^{D\over 2} 
\Theta\left(E_F- U({\bf r})\right) \; ,
$$
where $E_F$ is the Fermi energy.}  
\par
{\it Proof.} One finds the spatial distribution 
from the Eq. (5) by integrating over momenta the 
Eq. (3) (with the sign $+$). In particular, one has 
$$ 
n({\bf r})=\int {d^D{\bf p} \over (2\pi \hbar )^D} 
{1\over e^{\beta( {{\bf p}^2\over 2m} + U({\bf r}) -\mu)} + 1} = 
$$
$$
= {1\over (2\pi\hbar^2)^D} {D \pi^{D\over 2} \over \Gamma({D\over 2}+1)} 
\int_0^{\infty} dp 
{ p^{D-1} e^{\beta(\mu -U({\bf r}))} e^{-\beta( {p^2\over 2m})}
\over 
1 + e^{\beta(\mu - U({\bf r}))} e^{-\beta( {p^2\over 2m})} }  
\; , 
$$ 
where ${D \pi^{D\over 2}/\Gamma({D\over 2}+1)}$ 
is the volume of the D-dimensional unit sphere. 
Then, with the position $y^2 = \beta {p^2\over 2m}$ and 
using the Fermi function $f_{D\over 2}(z)$ with $z= 
e^{\beta(\mu - U({\bf r}))}$, one gets the 
finite temperature spatial distribution. 
Finally, one obtains the zero-temperature result  
by observing that 
$$ 
n({\bf r})=\int {d^D{\bf p} \over (2\pi \hbar )^D} 
\Theta\left(E_F - {{\bf p}^2\over 2m} - U({\bf r}) \right) = 
$$ 
$$
= {1\over (2\pi\hbar^2)^D} {D \pi^{D\over 2} \over \Gamma({D\over 2}+1)} 
\Theta\left(E_F - U({\bf r}) \right) 
\int_0^{\sqrt{2m(E_F-U({\bf r}))} } dp \; p^{D-1} = 
$$
$$ 
= {1\over (2\pi\hbar^2)^D} {D \pi^{D\over 2} \over \Gamma({D\over 2}+1)} 
\Theta\left(E_F - U({\bf r}) \right) {1\over D} 
\left( \sqrt{2m(E_F-U({\bf r})} \right)^D 
\; , 
$$ 
where the spatial distribution is taken from the Eq. (8). 
{\hfill $~\Box$} 

\vskip 0.5cm 

In the same way, but using the sign $-$ in Eq. (3) and 
the Bose function $g_{D\over 2}(z)$ with 
$z=e^{\beta(\mu - U({\bf r}))}$, 
one can easily prove also the following theorem. 

\vskip 0.5cm 

\par
THEOREM 1.2 
{\it For an ideal Bose gas in an external potential $U({\bf r})$, 
the finite temperature non-condensed spatial distribution is given by  
$$
n({\bf r})={1\over \lambda^D} g_{D\over 2}
\left(e^{\beta(\mu -U({\bf r}))}\right) \; ,
$$
where $\lambda = (2\pi \hbar^2\beta /m)^{1/2}$ is the 
thermal length and $\mu$ is the chemical potential.} 
{\hfill $~\Box$} 

\vskip 0.5cm

These two theorems are the generalization, 
of the formulas for ideal homogenous Fermi 
and Bose gases in a box of volume $V$ 
(for $D=3$ see Ref. 15). They show that, in the semiclassical limit, 
the non-homogenous results are obtained 
with the substitution $\mu \to \mu - U({\bf r})$, 
also called {\it local density approximation}. 
In particular, with $U({\bf r})=0$, from the 
previous theorems, one obtains 
the Fermi temperature $T_F$ and 
the Bose temperature $T_B$ for quantum gases 
in a rigid box, by imposing 
the normalization condition (4). The results are  
$$
E_F=kT_F= \left({2\pi\hbar^2\over m}\right)
\left[ \Gamma\left({D\over 2}+1\right) n \right]^{2/D}
$$ 
and 
$$
kT_B=\left({2\pi\hbar^2\over m}\right)
\left({n\over \zeta({D\over 2})}\right)^{2/D} 
\; ,
$$  
where $n=N/V$ is the homogenous density of particles 
(again, for $D=3$ see Ref. 15). 
\par 
In general, to find the momentum 
distribution, the Fermi temperature and the Bose temperature, 
it is necessary to specify the external potential. 
In many experiments with alkali-metal atoms, the 
external trap can be accurately modelled by a harmonic potential.$^{3}$ 
More generally, one can consider power-law potentials, 
which are important for studying 
the effects of adiabatic changes in the trap. 
The density of states of a quantum 
gas in the power-law potential $U({\bf r})=A\; r^n$
can be calculated from Eq. (11) and reads 
\beq 
\rho(\epsilon ) = 
\left({m\over 2\hbar^2}\right)^{D\over 2} 
\left({1\over A}\right)^{D\over \alpha} 
{\Gamma({D\over \alpha}+1) 
\over \Gamma({D\over 2}+1) 
\Gamma({D\over 2} +{D\over \alpha})} \epsilon^{{D\over 2} 
+{D\over \alpha} -1}  
\; . 
\eeq 
\par
We can now state two theorems about ideal Fermi and Bose gases 
in isotropic power-law potentials. 

\vskip 0.5cm

\par
THEOREM 2.1 {\it Let us consider an ideal Fermi gas 
in a power-law isotropic 
potential $U({\bf r})=A\; r^n$ with $r=|{\bf r}|=
(\sum_{i=1}^D x_i^2)^{1/2}$. 
The finite temperature momentum distribution is given by 
$$ 
n({\bf p})= {1 \over (2\hbar \sqrt{\pi})^D } 
{\Gamma({D\over n}+1) \over \Gamma({D\over 2}+1)} 
\left({1\over \beta A}\right)^{D\over n} 
f_{D\over n}\left(e^{\beta \left( \mu - {{\bf p}^2\over 2m} 
\right)}\right) \; . 
$$ 
The zero temperature momentum distribution is 
$$
n({\bf p})= {1 \over (2\hbar \sqrt{\pi})^D} 
\left({1\over A}\right)^{D\over n} 
\left(E_F -{{\bf p}^2\over 2m}\right)^{D\over n} 
\Theta\left(E_F -{{\bf p}^2\over 2m} \right) \; . 
$$
The Fermi energy $E_F$ and the Fermi temperature $T_F$ 
are given by 
$$
E_F = k T_F = \left[ 
\left({2 \hbar^2 \over m}\right)^{D\over 2} A^{D\over n} 
{\Gamma({D\over 2}+1)
\over \Gamma({D\over n}+1) } 
\Gamma\left({D\over 2}+{D\over n}+1\right) 
N \right]^{1\over {D\over 2}+{D\over n}} \; ,  
$$
where $N$ is the number of Fermions in the gas. }
\par
{\it Proof.} One finds the finite temperature 
momentum distribution from the Eq. (6) and by integrating 
over space coordinates the Eq. (3) (with the sign $+$). 
In particular, one has 
$$ 
n({\bf p})=\int {d^D{\bf r} \over (2\pi \hbar )^D} 
{1\over e^{\beta( {{\bf p}^2\over 2m} + A r^n -\mu)} + 1}  = 
$$
$$
= {1\over (2\pi\hbar^2)^D} 
{D \pi^{D\over 2} \over \Gamma({D\over 2}+1)} 
\int_0^{\infty} dr 
{ r^{D-1} e^{\beta(\mu - {p^2\over 2m})} e^{-\beta A r^n} 
\over 
1 + e^{\beta(\mu - {p^2\over 2m})} e^{-\beta A r^n} } 
\; , 
$$ 
where again ${D \pi^{D\over 2}/\Gamma({D\over 2}+1)}$ 
is the volume of the D-dimensional unit sphere. 
Setting $y^2 = {\beta A r^n}$ and 
using the definition of Fermi function $f_{D\over 2}(z)$ 
with $z=e^{\beta(\mu - {p^2\over 2m})}$, 
one finds the first formula of the theorem. 
The zero-temperature results are obtained by observing that 
$$ 
n({\bf p})=\int {d^D{\bf r} \over (2\pi \hbar )^D} 
\Theta\left(E_F - {{\bf p}^2\over 2m} - U({\bf r}) \right) = 
$$ 
$$
= {1\over (2\pi\hbar^2)^D} {D \pi^{D\over 2} \over \Gamma({D\over 2}+1)} 
\Theta\left(E_F - {p^2\over 2m}\right) 
\int_0^{A^{-{1\over n}}
(E_F - { {\bf p}^2 \over 2m})^{1/n}} dr \; r^{D-1}  = 
$$ 
$$
= {1\over (2\pi\hbar^2)^D} {D \pi^{D\over 2} \over \Gamma({D\over 2}+1)} 
\Theta\left(E_F - {p^2\over 2m}\right) {1\over D} 
\left( A^{-{1\over n}}
(E_F - { {\bf p}^2 \over 2m} )^{1/n} \right)^D 
\; , 
$$ 
where the momentum distribution is taken from the Eq. (9). 
The Fermi energy $E_F$ and the Fermi 
temperature $T_F$ are found from the normalization condition of the 
zero-temperature momentum distribution. Namely, one finds 
$$
N=\int d^D{\bf p} 
{1 \over (2\hbar \sqrt{\pi})^D} 
\left({1\over A}\right)^{D\over n} 
\left(E_F -{{\bf p}^2\over 2m}\right)^{D\over n} 
\Theta\left(E_F -{{\bf p}^2\over 2m} \right) = 
$$
$$
= {1 \over (2\hbar \sqrt{\pi})^D} 
\left({1\over A}\right)^{D\over n} \int_0^{\sqrt{2mE_F}} dp \; p^{D-1} 
\left(E_F -{p^2\over 2m}\right)^{D\over n} 
\; .
$$
Setting $x={p^2\over 2m}$ and observing$^{18,19}$ that 
$$
\int_0^{E_F}dx\; x^{{D\over 2}-1} (E_F-x)^{D\over n} = 
{ \Gamma({D\over 2}) \Gamma({D\over n}+1) \over 
  \Gamma\left({D\over 2}+{D\over n} +1\right) } 
E_F^{{D\over 2}+{D\over n}} 
\; ,
$$
one obtains 
$$ 
N = \left({m\over 2 \hbar^2}\right)^{D\over 2} 
\left({1\over A}\right)^{D\over n} 
{ \Gamma({D\over 2}) \Gamma({D\over n}+1) \over 
  \Gamma\left({D\over 2}+{D\over n} +1\right) } 
E_F^{{D\over 2}+{D\over n}} \; .
$$
Finally, by inverting this formula one gets the Fermi energy $E_F$. 
{\hfill $~\Box$} 

\vskip 0.5cm 

THEOREM 2.2 {\it Let us consider an ideal Bose gas 
in a power-law isotropic potential $U({\bf r})=A\; r^n$ with $r=|{\bf r}|
=(\sum_{i=1}^D x_i^2)^{1/2}$.  
The finite temperature non-condensed momentum distribution is given by 
$$ 
n({\bf p})= 
{1 \over (2\hbar \sqrt{\pi})^D } 
{\Gamma({D\over n}+1) \over \Gamma({D\over 2}+1)} 
\left({1\over \beta A}\right)^{D\over n} 
g_{D\over n}\left(e^{\beta \left( \mu - {{\bf p}^2\over 2m} 
\right)}\right) \; . 
$$ 
The Bose transition temperature $T_B$ reads 
$$
k T_B = \left[ \left({2 \hbar^2\over m}\right)^{D\over 2} 
A^{D\over n} 
{\Gamma({D\over 2}+1) \over \Gamma({D\over n}+1) } 
{1\over \zeta({D\over 2} + {D\over n})} N 
\right]^{1\over {D\over 2}+{D\over n}} 
$$ 
and the condensed fraction is 
$$
{N_0\over N} =1-\left({T\over T_B}\right)^{{D\over 2}+{D\over n}} 
\; , 
$$ 
where $N$ is the number of Bosons in the gas.}
\par
{\it Proof.} 
The finite temperature momentum distribution 
can be found by following the procedure 
used in the proof of the previous theorem: 
from the Eq. (6) and by integrating 
over space coordinates the Eq. (3) (but 
with the sign $-$). It follows that one must use 
the Bose function $g_{D\over 2}(z)$ 
with $z=e^{\beta(\mu - {p^2\over 2m})}$. 
At the BEC transition temperature $T_B$, 
the chemical potential $\mu$ is zero and 
at $\mu=0$ the number $N$ of particles 
can be analytically determined. One has 
$$ 
N= \int d^D{\bf p} 
{1 \over (2\hbar \sqrt{\pi})^D } 
{\Gamma({D\over n}+1) \over \Gamma({D\over 2}+1)} 
\left({1\over \beta A}\right)^{D\over n} 
g_{D\over n}\left(e^{-\beta {{\bf p}^2\over 2m}}\right) =  
$$
$$
= 
{1 \over (2\hbar \sqrt{\pi})^D } 
{\Gamma({D\over n}+1) \over \Gamma({D\over 2}+1)} 
{D \pi^{D\over 2} \over \Gamma({D\over 2}+1)} 
\left({1\over \beta A}\right)^{D\over n} 
\sum_{i=1}^{\infty} {1\over i^{D\over 2} } 
\int_0^{\infty} dp p^{D-1} e^{-i \beta {p^2\over 2m}} \; ,
$$
where the Bose function has been written as a power series 
(see Remark 1). Setting $x=i\beta{p^2\over 2m}$ 
and observing$^{18,19}$ that 
$$
\int_0^{\infty} dx\; x^{D\over 2} e^{-x} = 
\Gamma\left({D\over 2} + 1\right) 
\; ,
$$
and also 
$$ 
\sum_{i=0}^{\infty} {1\over i^{{D\over 2}+{D\over n}} } 
= \zeta\left({D\over 2} + {D\over n}\right) \; ,
$$ 
one obtains 
$$ 
N = \left( kT \right)^{{D\over 2} + {D\over n}} 
\left({m\over 2 \hbar^2}\right)^{D\over 2} 
\left({1\over A}\right)^{D\over n} 
{\Gamma({D\over n}+1) 
\zeta({D\over 2} + {D\over n}) 
\over \Gamma({D\over 2}+1) } \; . 
$$
By inverting the function $N=N(T)$ one finds 
the transition temperature $T_B$. 
Below $T_B$, a macroscopic number $N_0$ of particle occupies 
the single-particle ground-state of the system. It follows 
that the previous equation gives the number $N-N_0$ 
of non-condensed particles and the condensed fraction is 
$N_0/N=1-(T/T_B)^{{D/2}+{D/n}}$. {\hfill $~\Box$} 

\vskip 0.5cm 

\par 
This last theorem generalizes the BEC results obtained 
with $D=3$ by Bagnato, Pritchard and Kleppner.$^{17}$ 

\vskip 0.5cm 

It is important to observe that from the two previous theorems 
one easily derives the thermodynamic properties of quantum gases 
in harmonic traps and in a rigid box. 
In fact, by setting $n =2$ and $A=m\omega^2r^2/2$ one gets 
the formulas for the Bose and Fermi gases in a harmonic trap 
(in the case of a anisotropic harmonic potential, $\omega$ 
is the geometric average of the frequencies of the trap). 
The results for a rigid box are instead obtained by letting 
${D\over n}\to 0$, where the density of particles 
per unit length is given by $N/\Omega_D$ and 
$\Omega_D= D\pi^{D\over 2}/
\Gamma({D\over 2}+1)$ is the volume of the D-dimensional unit sphere. 
\par 
Finally, one notes that in the formula of the 
BEC transition temperature $T_B$ 
it appears the function $\zeta({D\over 2}+{D\over n})$. 
Because $\zeta(x)< \infty$ for $x>1$ but 
$\zeta(1)=\infty$,$^{17,18}$ one easily deduces 
the following corollary. 

\vskip 0.5cm 

COROLLARY 1. {\it Let us consider an ideal Bose gas 
in a power-law isotropic potential $U({\bf r})=A\; r^n$ 
with $r=|{\bf r}|=(\sum_{i=1}^D x_i^2)^{1/2}$. 
BEC is possible if and only if the following condition 
is satisfied  
$$
{D\over 2}+{D\over n} > 1\; ,
$$
where $D$ is the space dimension and $n$ is the exponent of the 
confining power-law potential. 
{\hfill $~\Box$} } 

\vskip 0.5cm 

This is a remarkable inequality. For example, 
for $D=2$ one finds the familiar result that 
there is no BEC in a homogenous gas 
(${D\over n} \to 0$) but BEC is possible in a harmonic trap 
($n =2$). Moreover, one obtains that 
for $D=1$ BEC is possible with $1 < n < 2$. 

\section{Conclusions}

By using the grand canonical ensemble of statistical 
mechanics and the semiclassical approximation, 
we have derived some thermodynamic properties of 
ideal quantum gases in a generic isotropic power-law 
confining external potential. 
We have calculated the density of states, spatial and momentum 
distributions and obtained analytical formulas for 
the Fermi energy, the BEC transition temperature 
and the Bose condensed fraction. 
Note that nowadays the spatial and momentum density profiles 
are quantities easily experimentally measured. 
We have also shown that BEC in an isotropic power-law 
potential is possible if and only if 
${D/2}+{D/n}$, where $D$ is 
the space dimension and $n$ is 
the exponent of the confining power-law potential. 
\par
The present investigation is the starting point for 
future analyses of interacting quantum gases in D-dimensional 
space and generic trapping potential. 

\newpage 

\section*{References}

\begin{description}

\item{\ $^{1}$} M.H. Anderson, J.R. Ensher, M.R. Matthews, 
C.E. Wieman and E.A. Cornell, 
``Observation of Bose-Einstein condensation in a dilute atomic vapor'', 
{\it Science} {\bf 269}, 198-201 (1995). 

\item{\ $^{2}$} B. DeMarco and D.S. Jin, 
``Onset of Fermi degeneracy in a trapped atomic gas'', 
{\it Science} {\bf 285}, 1703-1706 (1999). 

\item{\ $^{3}$} F. Dalfovo, S. Giorgini, L.P. Pitaevskii 
and S. Stringari,  
``Theory of Bose-Einstein condensation in trapped gases'', 
{\it Rev. Mod. Phys.} {\bf 71}, 463-512 (1999). 

\item{\ $^{4}$} S. Inouye, M.R. Andrews, J. Stenger, H.J. Meisner, 
D.M. Stamper-Kurn and W. Ketterle, 
``Observation of Feshback resonances in a Bose-Einstein condensate'', 
{\it Nature} {\bf 392}, 151-154 (1998). 

\item{\ $^{5}$} L. Salasnich, 
``The role of dimensionality in the stability of 
a confined condensed Bose gas'', 
{\it Mod. Phys. Lett. B} {\bf 11}, 1249-1254 (1997).   

\item{\ $^{6}$} L. Salasnich, 
``Note on the role of 
dimensionality in the stability of 
a confined condensed Bose gas: reply to a comment'', 
{\it Mod. Phys. Lett. B} {\bf 12}, 649-651 (1998). 

\item{\ $^{7}$} E. Cerboneschi, R. Mannella, E. Arimondo 
and L. Salasnich, ``Oscillation frequencies 
for a Bose condensate confined in a 
triaxially anisotropic magnetic trap'', 
{\it Phys. Lett. A} {\bf 249}, 245-500 (1998). 

\item{\ $^{8}$} L. Salasnich, 
``Self-trapping, quantum tunneling and decay rates 
for a Bose gas with attractive nonlocal interaction'', 
{\it Phys. Rev. A} {\bf 61}, 015601 (2000). 

\item{\ $^{9}$} L. Salasnich, 
``Time-dependent variational approach to Bose-Einstein condensation'', 
{\it Int. J. Mod. Phys. B} {\bf 14}, 1-11 (2000). 

\item{\ $^{10}$} L. Salasnich, 
``Resonances and chaos in the collective oscillations of a trapped 
Bose condensate'', 
{\it Phys. Lett. A}, vol. {\bf 266}, 187-192 (2000). 

\item{\ $^{11}$} L. Salasnich, A. Parola and L. Reatto, 
``Bose condensation in a toroidal trap: ground state and vortices'', 
{\it Phys. Rev. A} {\bf 59}, 2990-2995 (1999). 

\item{\ $^{12}$} L. Salasnich, A. Parola and L. Reatto, 
``Bose condensate in a double-well trap: ground state and 
elementary excitations'', {\it Phys. Rev. A} {\bf 60}, 4171-4174 (1999). 

\item{\ $^{13}$} B. Pozzi, L. Salasnich, A. Parola and L. Reatto, 
``Thermodynamics of trapped Bose condensate with 
negative scattering length'', 
{\it J. Low Temp. Phys.} {\bf 113}, 57-77 (2000). 

\item{\ $^{14}$} L. Salasnich, 
``BEC in nonextensive statistical mechanics'', 
{\it Int. J. Mod. Phys. B} {\bf 14}, 405-410 (2000). 

\item{\ $^{15}$} K. Huang, {\it Statistical Mechanics} 
(John Wiley, New York, 1987). 

\item{\ $^{16}$} A.M.O. De Almeida, 
{\it Hamiltonian Systems: Chaos and Quantization} (Cambridge 
University Press, Cambridge, 1988). 

\item{\ $^{17}$} V. Bagnato, D.E. Pritchard and D. Kleppner, 
``Bose-Einstein condensation in an external potential'', 
{\it Phys. Rev. A} {\bf 35}, 4354-4358 (1987). 

\item{\ $^{18}$} M. Abramowitz and I.A. Stegun, 
{\it Handbook of Mathematical Functions}, 
(Dover, New York 1965). 

\item{\ $^{19}$} I.S. Gradshteyn and I.M. Ryzhik 
{\it Table of Integrals, Series, and Products} 
(Academic Press, Orlando, 1980). 

\end{description}

\end{document}